\documentclass{aa}

\usepackage{graphicx}
\usepackage{natbib}
\usepackage[varg]{txfonts}
\usepackage{xcolor}
\definecolor{xlinkcolor}{rgb}{0, 0, 1}
\usepackage[bookmarks=true,
pdfnewwindow=true,
colorlinks=true,
linkcolor=xlinkcolor,
citecolor=xlinkcolor,
filecolor=xlinkcolor,
urlcolor=xlinkcolor,
final=true]{hyperref}
\begin{document}

\title{Revisiting the abundance pattern and charge-exchange emission in the M82 centre}
\subtitle{}
\titlerunning{Abundance pattern and charge-exchange emission in M82}

\author{
K.~Fukushima\inst{\ref{inst:tus}}
\href{http://orcid.org/0000-0001-8055-7113}{\includegraphics[width=9pt]{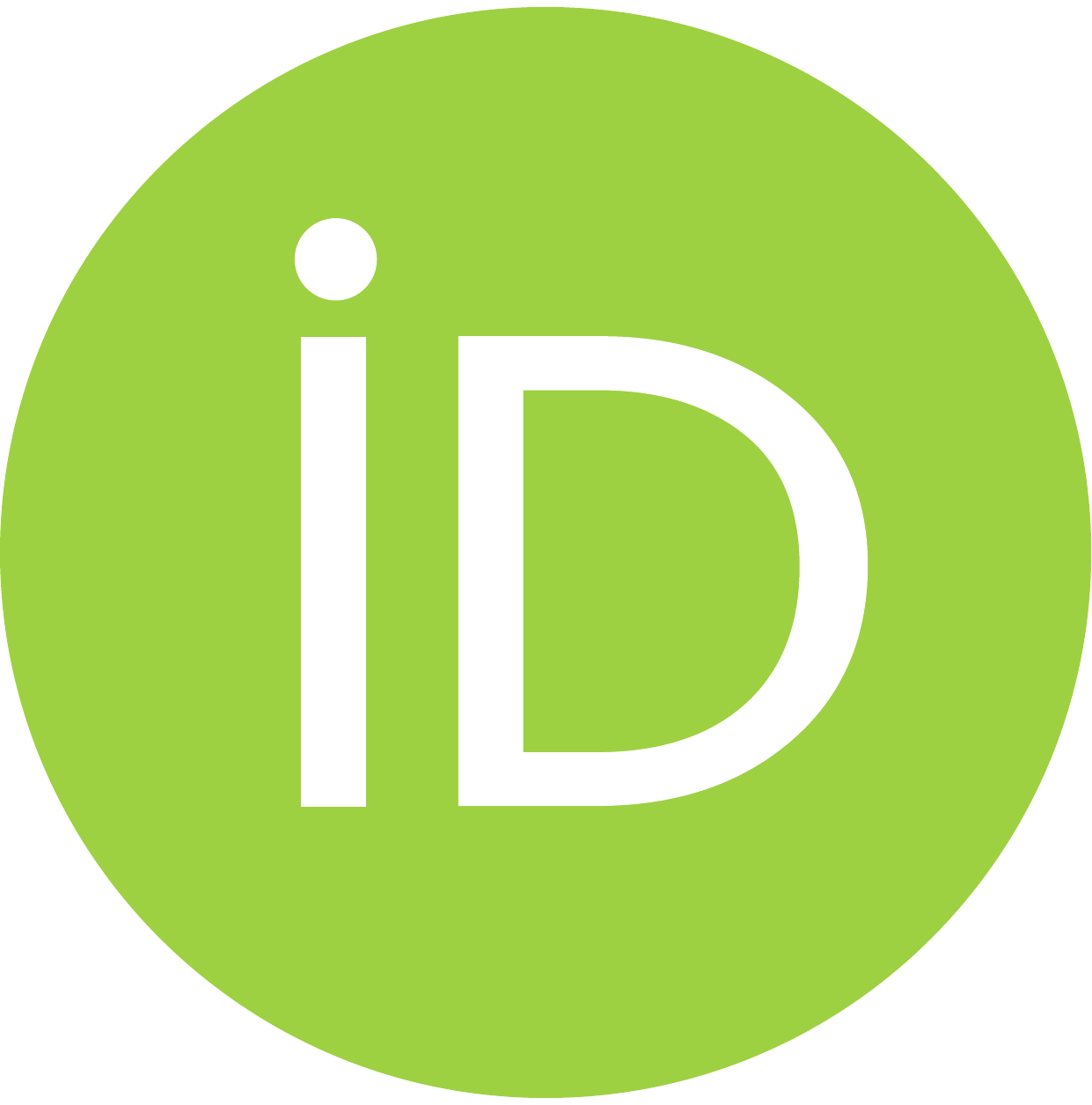}}
\and S.~B.~Kobayashi\inst{\ref{inst:tus}}
\href{http://orcid.org/0000-0001-7773-9266}{\includegraphics[width=9pt]{logo-orcid.pdf}}
\and K.~Matsushita\inst{\ref{inst:tus}}
\href{http://orcid.org/0000-0003-2907-0902}{\includegraphics[width=9pt]{logo-orcid.pdf}}
}
\authorrunning{K.~Fukushima et al.}

\institute{{Department of Physics, Tokyo University of Science, 1-3 Kagurazaka, Shinjuku-ku, 
Tokyo 162-8601, Japan\label{inst:tus}}
\\
\email{kxfukushima@gmail.com} (K.F.)
}

\date{Received / Accepted }


\abstract
{
The interstellar medium (ISM) in starburst galaxies contains plenty of chemical elements
synthesised by core-collapse supernova explosions.
By measuring the abundances of these metals, we can study
the chemical enrichment within galaxies
and the transportation of metals into circumgalactic environments through powerful outflows.
}
{We perform the spectral analysis of the X-ray emissions from the M82 core
using the Reflection Grating Spectrometer (RGS) onboard \textit{XMM-Newton}
to accurately estimate the metal abundances in the ISM.
}
{
We analyse over 300\,ks of RGS data observed with fourteen position angles,
covering an 80\,arcsec cross-dispersion width. 
We employ multi-temperature thermal plasma components
in collisional ionisation equilibrium (CIE) to reproduce the observed spectra,
each exhibiting different spatial broadenings.
}
{The \ion{O}{VII} band CCD image shows a broader distribution
compared to those for \ion{O}{VIII} and Fe-L bands.
The \ion{O}{VIII} line profiles have a prominent double-peaked structure,
corresponding to the northward and southward outflows.
The \ion{O}{VII} triplet feature exhibits marginal peaks, and a single CIE component,
convolved with the \ion{O}{VII} band image, approximately reproduces the spectral shape.
Combining a CIE model with a charge-exchange emission model
also successfully reproduces the \ion{O}{VII} line profiles.
However, the ratio of these two components varies significantly
with the observed position angles, which is physically implausible.
Spectral fitting of the broadband spectra suggests a multi-temperature phase in the ISM,
approximated by three components at 0.1, 0.4, and 0.7\,keV.
Notably, the 0.1\,keV component exhibits a broader distribution than the 0.4 and 0.7\,keV plasmas.
The derived abundance pattern shows super-solar N/O,
solar Ne/O and Mg/O, and half-solar Fe/O ratios.
These results indicate the chemical enrichments by core-collapse supernovae in starburst galaxies.
}
{}


\maketitle
%

\section{Introduction}
\label{sec:intro}

Core-collapse supernovae (CCSNe) from massive star progenitors
play a crucial role in the evolution of galaxies.
They are primary sources of light $\alpha$-elements such as O, Ne, and Mg \citep[e.g.,][]{Nomoto13}.
Additionally, they heat the interstellar medium (ISM) to X-ray-emitting temperatures.
In starburst galaxies, in particular regions of active star formation,
CCSNe can induce powerful winds from the galactic disc.
These winds, also known as outflows, are essential in the enrichment
of the circumgalactic medium that connects the ISM to the local Universe
\citep[][for a comprehensive review]{Orlitova20}.
Consequently, X-ray observations of starburst galaxies offer valuable insights into
the processes supplying energies and metals into both the ISM and circumgalactic spaces.

In recent years, numerous X-ray observations have focused on the extended X-ray-emitting ISM
in starburst galaxies \citep[e.g.,][]{Strickland04a, Yamasaki09, Konami12, Mitsuishi13, Yang20}.
The M82 galaxy, a well-established starburst system located
at a 3.53\,Mpc distance from our solar system
($1' \sim$\,1\,kpc thereat, \citealt{Karachentsev04}),
has been a key object among these studies.
The nearly edge-on appearance makes this galaxy an ideal laboratory
for studying biconical outflows that erupt from the central region of the galaxy
\citep[e.g.,][]{Strickland09, Zhang18}.
These outflows exhibit a multiphase structure,
which is observed in the infrared, H$\alpha$, and X-ray windows
\citep[][]{Strickland97, Lehnert99, RS02, Engelbracht06, Zhang14b, Lopez20}.
\citet{SH07} found diffuse Fe He$\alpha$ emission in the central regions, 
suggesting the presence of gas at temperatures of several kiloelectronbolts.
Additionally, strong emission lines of elements such as O, Mg, Si, S, and Fe
have been detected from the core to the wind regions of the galaxy
\citep[e.g.,][]{Tsuru97, Tsuru07, Ranalli08, Konami11}.
In the central regions, these elements show an unusual abundance pattern:
super-solar ratios of Ne/O and Mg/O ($> 1.5$\,solar) with super-solar O/Fe ratios
have been reported \citep[e.g.,][]{Tsuru97, Ranalli08, Konami11, Zhang14b}.
This contradicts the pre-existing nucleosynthesis models of CCSN \citep[e.g.,][]{Nomoto13}.
In contrast, the outflowing gas outside the disc shows an abundance pattern
consistent with CCSN predictions\citep[][]{Tsuru07, Konami11}.

An additional exciting feature in M82 is the presence of charge exchange (CX) X-ray emission. 
In this process, one or more electrons are transferred from a cold neutral atom to a hot ion.
Subsequently, the cascade relaxation of such a captured electron from a highly excited
energy state to the ground state of the ion result in the X-ray line emission
\citep[e.g.,][]{Cravens02, Sibeck18, GS23}.
A notable feature of the CX process is present in the \ion{O}{VII} triplet structure:
the enhanced forbidden line at 0.56 keV compared to the resonance line at 0.57 keV.
In starburst galaxies, it is expected to observe CX emissions
at the fronts of hot gas flows colliding with cold matter.
\citet{Liu12a} analysed the data from the Reflection Grating Spectrometer (RGS)
on board \textit{XMM-Newton} and reported this enhancement in the \ion{O}{VII} triplet
in several starburst galaxies, suggesting a significant CX contribution
in addition to plasma in a collisional ionisation equilibrium (CIE) state.
\citet{Konami11} and \citet{Lopez20} included the CX component
in their analysis of the CCD spectra from M82,
although their spectral resolution was limited to resolve the \ion{O}{VII} triplet sufficiently.
\citet{Zhang14b} used the RGS data from an \textit{XMM-Newton} observation
of M82 and estimated that the interaction area between the neutral and ionised matter
is significantly larger than the geometry surface area of the galaxy.

The RGS on board \textit{XMM-Newton} is effective for our study, 
offering good spatial and spectral resolution within the $\sim$\,0.5--2\,keV band.
This makes it particularly useful for measuring $\alpha$-element abundances
and estimating the CX contribution.
However, any line structures in the dispersed spectra
are practically broadened when the source extends $\gtrsim 0.8$\,arcmin,
leading to decreased resolving power for extended objects
such as galaxies \citep[][for a recent review]{Mao23}.
Therefore, careful treatment is required when analysing the RGS spectra of galaxies,
especially for the broadening effect on each emission line
\citep[e.g.,][]{Chen18, Yang20, Fukushima23a}.
In this study, we used deep 300\,ks RGS data obtained with different position angles to model
the spectra of the central region of M82 
to determine its temperature structure, the CX contribution, and the abundance pattern.
The paper is structured as follows. Section\,\ref{sec:observation} summarises
the \textit{XMM-Newton} observations and data reduction.
In Sect.\,\ref{sec:lpro}, we outline the line-broadening effect intrinsic to the RGS analysis.
Section\,\ref{sec:result} presents the general prescription for our fitting and the results.
We will discuss and interpret our findings in Sect.\,\ref{sec:discussion}.
We assume cosmological parameters as $H_0 = 70$\,km\,s$^{-1}$\,Mpc$^{-1}$, 
$\Omega_\textup{m} = 0.3$, and $\Omega_{\Lambda} = 0.7$.
The proto-solar abundance of \citet{Lodders09} is adopted to estimate element abundances.

\section{Observations and data reduction}
\label{sec:observation}

\begin{table}
\centering
\caption{The RGS observations analysed in this work. \label{tab:observation}}
\begin{tabular}{@{}c@{\:\:\:}c@{\:\:\:}r@{\:\:\:}r}\\ \hline\hline
ObsID & Date & Exposure\tablefootmark{a} & Position angle\tablefootmark{b} \\
 & & (ks) & (deg) \\ \hline
0112290201 & 2001 May 06 & 19.2 & 293 \\
0206080101\tablefootmark{c} & 2004 Apr 21 & 55.3 & 319 \\
0560590101 & 2008 Oct 03 & 27.8 & 138 \\
0560181301 & 2009 Apr 03 & 22.9 & 316 \\
0560590201 & 2009 Apr 17 & 15.1 & 316 \\
0560590301 & 2009 Apr 29 & 21.4 & 296 \\
0657800101 & 2011 Mar 18 & 22.8 & 330 \\
0657801701 & 2011 Apr 11 & 21.4 & 312 \\
0657801901 & 2011 Apr 29 & 11.6 & 296 \\
0657802101 & 2011 Sep 24 & 12.4 & 147 \\
0657802301 & 2011 Nov 21 & 9.2 & 100 \\
0870940101 & 2021 Apr 06 & 28.6 & 313 \\
0891060101 & 2021 Oct 17 & 26.8 & 127 \\
0891060401 & 2022 Apr 06 & 32.1 & 313 \\ \hline
\end{tabular}
\tablefoot{
\tablefoottext{a}{Half the sum of the RGS1 and RGS2 exposures
after the deflaring procedure.}
\tablefoottext{b}{The angle of the RGS dispersion axis measured eastwards
from the celestial north.}
\tablefoottext{c}{The data set analysed in \citet{Zhang14b}.}
}
\end{table}

We analysed the archival observation data of the M82 centre
with RGS installed on \textit{XMM-Newton}.
Table\,\ref{tab:observation} summarises the observations used in this work.
The M82 nucleus \citep[][]{Lester90} is located near the centre of the MOS detecter
for these observations.
The RGS data were processed using the \texttt{rgsproc} task
wrapped in the \textit{XMM-Newton} Science Analysis System (\textsc{sas}) version 18.0.0.
We obtained light curves from the CCD9 on the RGS focal plane \citep{denHerder01};
and next, the mean count rate $\mu$ and standard deviation $\sigma$ were derived
by fitting these curves with a Gaussian.
In order to discard background soft-proton flares,
the count rate threshold as $\mu \pm 2\sigma$ was applied to uncleaned events.
Following the \textsc{sas} team, other standard screening criteria were employed.
The cleaned exposure times of each observation are in Table\,\ref{tab:observation}.

\section{Imaging analysis}
\label{sec:lpro}

\begin{figure*}
\centering
\includegraphics[width=0.85\linewidth]{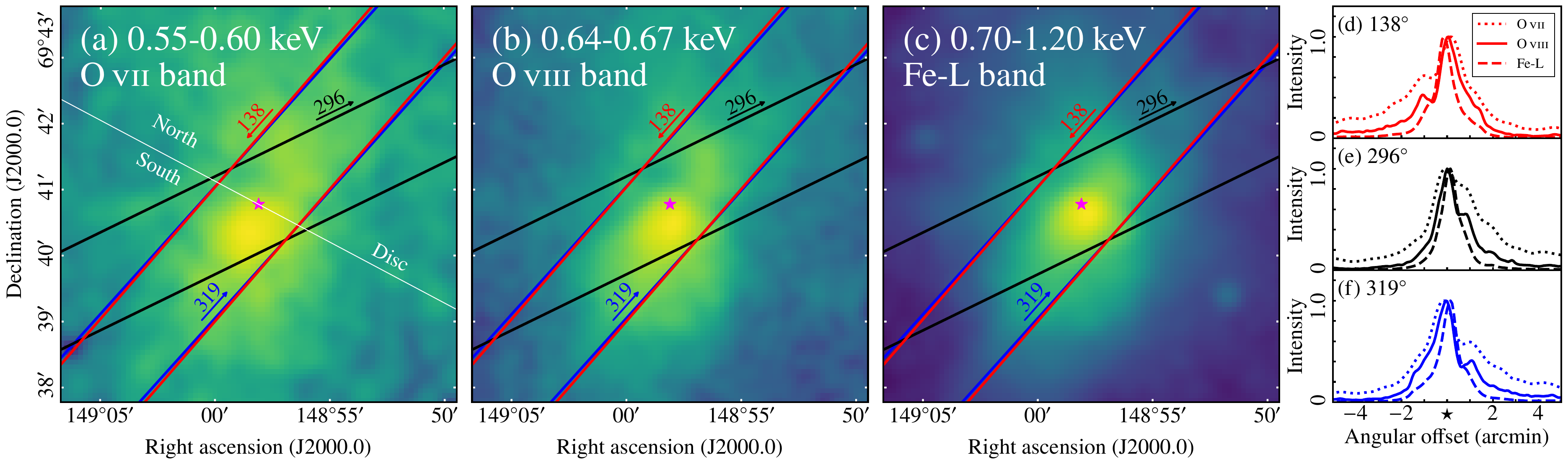}
\caption{(a) Mosaiced X-ray count image of the M82 centre in the \ion{O}{VII}
(0.55--0.60\,keV) band by CCDs.
The representative three RGS position angles are overlaid with the 80\,arcsec cross-dispersion widths.
The arrows indicate each dispersion direction
in which the photons are scattered into the higher energy side.
The star marker and the solid line passing through it
indicate the M82 centre and disc \citep[][]{Lester90, Mayya05}.
The continuum flux estimated using the 0.45--0.55\,keV band is subtracted from the raw count image.
(b, c) Same as (a), but in the \ion{O}{VIII}~Ly$\alpha$ (0.64--0.67\,keV)
and Fe-L (0.70--1.20\,keV) bands.
(d) Normalized count profiles projected on to the 138$^\circ$ dispersion axis
in the 0.55--0.60\,keV (\ion{O}{VII}), 0.64--0.67\,keV (\ion{O}{VIII}),
and 0.70--1.20\,keV (dominated by Fe-L lines) bands.
The angular offset is measured in reference to the star marker in (a),
where the positive side of the offset axis is the higher energy part in diffracted spectra.
(e, f) Same as (d), but along the dispersion axes of 296$^\circ$ and 319$^\circ$, respectively.
\label{fig:image}}
\end{figure*}

In Figs.\,\ref{fig:image}(a), (b), and (c), we show X-ray images of the M82 centre
for the \ion{O}{VII} (Fig.\,\ref{fig:image}(a), 0.55--0.60\,keV),
\ion{O}{VIII}~Ly$\alpha$ (Fig.\,\ref{fig:image}(b), 0.64--0.67\,keV),
and Fe-L (Fig.\,\ref{fig:image}(c), 0.70--1.20\,keV) bands.
Avoiding relatively large noises in the \ion{O}{VII} band image,
we subtract the continuum counts from the raw events.
Point-like sources are not discarded.
In these images, the outflows extending from the disc
towards the southeast and northwest are visible,
especially in Figs.\,\ref{fig:image}(a) and (b).
The dispersion directions for the three representative position angles are overlayed.
The 80\,arcsec cross-dispersion width covers a significant part of the two outflow regions.
Incorporating data from different position angles helps resolve the degeneracy
between energy and spatial position along the dispersion angle,
allowing for a more detailed analysis of the outflow structure.

The X-ray count profiles within an 80\,arcsec cross-dispersion width
along each dispersion axes are given in Figs.\,\ref{fig:image}(d), (e), and (f).
These profiles show significant differences across the three bandpasses.
In particular, the Fe-L line band is characterised by a centrally peaked and narrow 
distribution on a spatial scale of 1--2\,arcmin.
In contrast, the \ion{O}{VII} band exhibits the most broadened profiles.
In both the \ion{O}{VII} and \ion{O}{VIII} profiles, a dip is observed
at $\sim$\,1\,arcmin offset northward from the galaxy centre.
The two peaks beside the dip correspond to the outflows towards the southeast and northeast.
This bimodal line profile in the O band, along the centrally-peaked distribution
in the Fe-L line band, has been reported by \citet{Zhang14b}.
The appearances of this double-peak structure, especially in the \ion{O}{VIII} band profile,
depends on the dispersion direction (Figs.\,\ref{fig:image}(d) and (f)).

\section{Spectral analysis}
\label{sec:result}

\subsection{General prescription}
\label{subsec:fitting}

The RGS spectra along the dispersion direction can be used
within a 5\, arcmin cross-dispersion width centred on the MOS detector \citep[e.g.,][]{Zhang19, Narita23}.
After filtering the cleaned event files with \texttt{XDSP\_CORR} columns using \texttt{evselect},
we extracted first-order RGS spectra from the 80\,arcsec cross-dispersion width
centred on the M82 nucleus by the \texttt{rgsspectrum} task.
This method takes advantage of determining accurately the rectangular cross-dispersion limits
\footnote{Alternatively, one can use the \texttt{rgsregions} task to do this
if specifying a rectangle that is roughly symmetrical about the detector centre.
Otherwise, users are strongly urged to check the generated RMFs and effective areas.}
compared to the standard RGS analysis with \texttt{xpsf}- options of \texttt{rgsproc}, for instance
\citep[e.g.,][]{Fukushima22, Fukushima23a}.
The cross-dispersion widths and dispersion directions for three representative position angles
are plotted in Figs.\,\ref{fig:image}(a), (b), and (c).
The RGS1 and RGS2 spectra and response matrices are co-added through the \texttt{rgscombine} script.

The RGS instrument is primarily designed for observing point-like sources;
therefore, the response matrices generated by the standard \textsc{sas} method
are unsuitable for analysing diffuse emissions from various extended sources,
such as supernova remnants, galaxies, and clusters
\citep[e.g.,][]{Chen18, Tateishi21, Fukushima23a}.
Line broadening in a first-order RGS spectrum is observed as $\Delta E = 0.138 \Delta \theta$,
where $\Delta E$ and $\Delta \theta$ represent the line broadening and the spatial extent
in the wavelength (\AA) and dispersion (arcmin) axes, respectively.
To fit the RGS spectra of these extended sources,
it is necessary to convolve the response matrix with the projected surface-brightness profile
along the dispersion direction \citep[e.g.,][]{TO04, Mao23}.
Moreover, variations in spatial broadening across different energy bands
necessitate applying different convolution scales for distinct spectral components.
The differences in the spatial distribution of the \ion{O}{VII}, \ion{O}{VIII},
and Fe-L band images (Section\,\ref{sec:lpro})
imply the presence of multiple thermal CIE components with different spatial scales.
\citet{Zhang14b} carefully analysed the RGS spectra of M82,
adopting two spectral broadenings: bimodal soft emission and centrally-peaked hard emission.
In the subsequent part, we further test different spatial broadenings
for the spectra components emitting \ion{O}{VII} and \ion{O}{VIII} lines.

We use the \textsc{xspec} package version 12.10.1f \citep[][]{Arnaud96}
but with the revised \textsc{AtomDB} version 3.0.9 \citep[][]{Smith01, Foster12}
that provides information on the line and continuum emissions
at two hundred and one temperatures from $8.6 \times 10^{-4}$ to $86$\,keV.
This updated \textsc{AtomDB} is implemented standardly in \textsc{xspec} version 12.11.0k
or later \footnote{\url{http://www.atomdb.org}}.
The C-statistic method \citep[][]{Cash79} is adopted in our spectral fitting
to estimate the spectral parameters and the error ranges thereof without bias \citep[][]{Kaastra17}.
Each spectrum is re-binned to have a minimum of 1\,count per spectral bin.

\subsection{Spectral models}

In our RGS spectral analysis, we will reproduce the diffuse ISM emission
(denoted by $\textup{ISM}_\textup{M82}$).
Two individual absorptions modify the $\textup{ISM}_\textup{M82}$ component.
We assume $6.7 \times 10^{20}$\,cm$^{-2}$ for the Galactic extinction \citep[][]{Willingale13}.
The intrinsic absorption of M82 for diffuse emissions
is typically (1--3)\,$\times 10^{21}$\,cm$^{-2}$ \citep[e.g.,][]{Konami11, Zhang14b};
thus, we use the fixed value $2 \times 10^{21}$\,cm$^{-2}$.
The photoelectric absorption cross-sections are retrieved from \citet{Verner96}.
Letting these absorption parameters free does not
significantly change the results we demonstrate below.
We do not include diffuse astrophysical background emissions such as
the cosmic X-ray background or Milky Way halo component in our spectral fitting
since they contribute to the observed RGS spectra just as flat components.
A power-law component is introduced to account for
increasing flux at low energies below 0.5\,keV.
We use another power-law component with a 1.6 photon index \citep[][]{Ranalli08}
subjected to $2 \times 10^{22}$\,cm$^{-2}$ extinction
to take the point source emission into account.

The complete model applied to the observed spectra is represented as
$\texttt{phabs}_\textup{Gal} \times (\texttt{phabs}_\textup{M82} \times \textup{ISM}_\textup{M82}
+ \texttt{phabs}_\textup{PS} \times \texttt{powerlaw}_\textup{PS}) + \texttt{powerlaw}_\textup{BKG}$.
To account for the line broadening effect on $\textup{ISM}_\textup{M82}$
described in Sects.\,\ref{sec:intro} and \ref{sec:lpro},
we use the \textsc{xspec} model \texttt{rgsxsrc} with mosaiced CCD (MOS+pn) images in specific bands.
We first perform spectral fits in two local bands: 
the \ion{O}{VIII} band (0.60--0.77\,keV, Sect.\,\ref{subsec:oviii})
and \ion{O}{VII} band (0.45--0.62\,keV, Sect.\,\ref{subsec:ovii}),
each convolved with the respective images.
Then, we fit the broadband spectra
covering the full RGS energy range of 0.45--1.75\,keV
(hereafter broadband; Sect.\,\ref{subsec:broad}),
using the image of this band, which closely resembles the Fe-L band image.
The spatial distribution of each emission line
may not precisely match the image of its corresponding energy band,
as these bands also include continuum emissions.
The image convolution approach in this study is an approximation, as well as similar works,
and the resulting systematic uncertainties will be discussed in further sections.

\subsection{Results of \ion{O}{VIII} Ly$\alpha$ line}
\label{subsec:oviii}

\begin{figure*}
\centering
\includegraphics[width=0.85\linewidth]{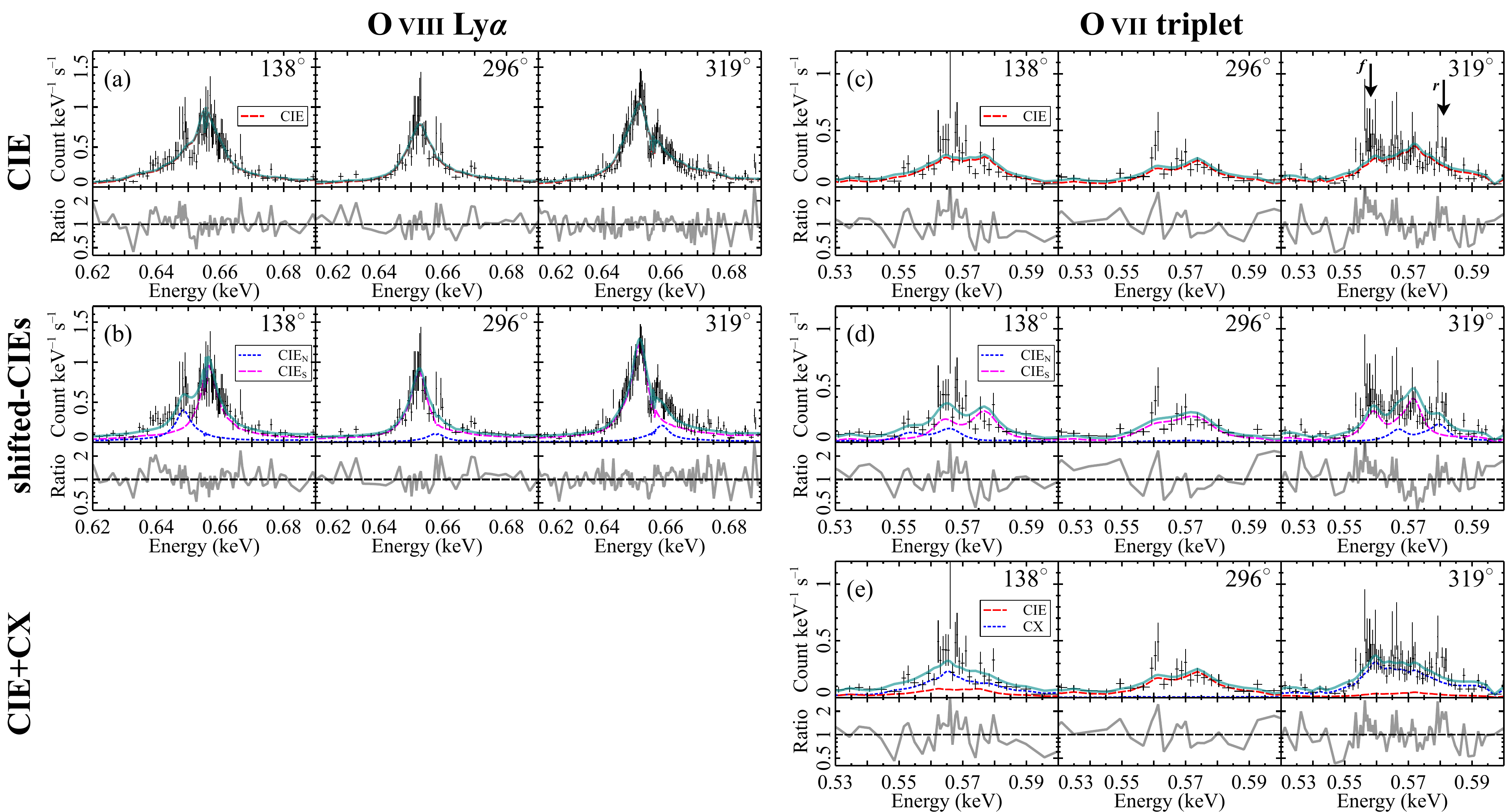}
\caption{The RGS spectra of the M82 centre
in the \ion{O}{VIII} (a, b) and \ion{O}{VII} (c, d, e) bands for the three representative position angles.
The best-fitting models and the data-to-model ratios for the three methods are plotted:
the CIE modelling convolved with local band images (the first row),
the shifted-CIEs modelling convolved by the broadband image (the second row),
and the CIE+CX modelling convolved with the \ion{O}{VII} band image
(the third row, only for the \ion{O}{VII} lines).
The positions of the \ion{O}{VII} forbidden and resonance lines
are labelled in the 319$^\circ$ panel of (c).
\label{fig:lspec}}
\end{figure*}

\begin{figure*}
\centering
\includegraphics[width=0.85\linewidth]{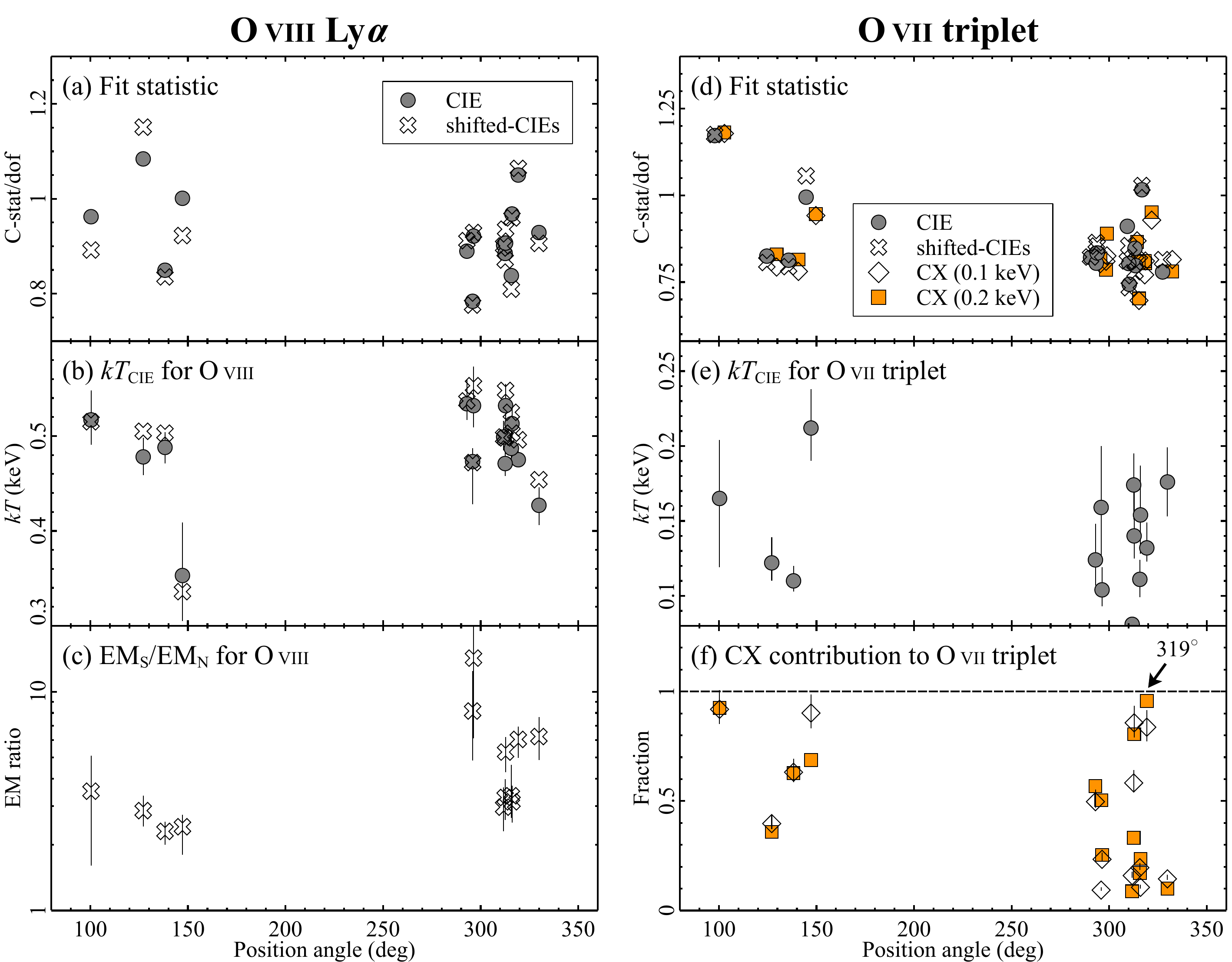}
\caption{(a) C-stats/dof values from the CIE and shifted-CIEs modellings
on the 0.60--0.77\,keV band (\ion{O}{VIII} and \ion{Fe}{XVII})
plotted against the RGS position angle.
(b) Temperatures derived from the two modellings.
(c) The $\textup{EM}_\textup{S}/\textup{EM}_\textup{N}$ profile
obtained from the shifted-CIEs method.
(d) Same as (a), but on the 0.45--0.62\,keV band fitting (\ion{N}{VII} and \ion{O}{VII}).
The results are from the CIE, shifted-CIEs (0.14\,keV), and CIE+CX modellings.
(e) Temperatures derived from the CIE modelling.
(f) Flux contribution of CX components to the \ion{O}{VII} line.
\label{fig:ofit}}
\end{figure*}

First, we test the thermal plasma component in collisional ionisation equilibrium (CIE)
convolved with the \ion{O}{VIII} band image:
$\textup{ISM}_\textup{M82} = \texttt{vapec}$ (the CIE modelling).
We ignore the point source contribution due to quite a strong intrinsic absorption
for the point sources at the M82 centre
($\gtrsim 1 \times 10^{22}$\,cm$^{-2}$, e.g., \citealt{Brightman16}).
Despite focusing on local line structures, in the spectral fittings,
we use a wider energy range of 0.60--0.77\,keV,
including the \ion{Fe}{XVII} line around 0.73\,keV.
This approach enabled more robust measurement of global parameters such as $kT$;
restricting to ``local fits'' lead to poorly constrained values \citep[e.g.,][]{Lakhchaura19}.
The combined spectrum of each observation
is fitted with free parameters of $kT$, emission measure (EM),
and the O and Fe abundances.

The CIE modelling approximately reproduces the observed spectra (Fig.\,\ref{fig:lspec}(a)),
achieving good C-statistics/dof values $\lesssim 1.2$ (the $x$-axes of Fig.\,\ref{fig:ofit}(a)).
However, a spectral sub-peak corresponding to the northward wind region
at $\sim$0.65 keV and $\sim$0.66 keV for the position angle
of 138$^\circ$ and 319$^\circ$, respectively,
appears narrower than the spectral model convolved with the \ion{O}{VIII} band image.
The derived $kT_\textup{CIE}$ values are plotted
against the RGS position angle in Fig.\,\ref{fig:ofit}(b),
yielding a uniform $kT_\textup{CIE}$ with a median of 0.49\,keV
and a 16--84th percentile range of 0.47--0.53\,keV.

Figure\,\ref{fig:ofit} shows that the \ion{O}{VIII}~Ly$\alpha$ consists of
a main peak and sub peak $\sim$\,5 eV away from it,
whose intensities depend on the RGS dispersion direction.
The 5\,eV separation between the two peaks at $\sim$\,0.66 keV
corresponds to an angular diameter of 1\,arcmin along the dispersion direction (Fig.\,\ref{fig:image}(b)).
To reproduce this double-peak structure,
we then try two spatially distinct components with varying redshift parameters
to represent different spatial positions:
$\textup{ISM}_\textup{M82} = \texttt{vapec}_\textup{N} + \texttt{vapec}_\textup{S}$
(the shifted-CIEs modelling).
Here, subscripts ``N'' and ``S'' represent the northeast and southwest emission peaks, respectively.
The temperature and metal abundances of the two components are assumed to have the same values.
Each CIE component is convolved with the broadband image (0.45--1.75\,keV)
as the extent of each peak is similar to that of the Fe-L band (Figs.\,\ref{fig:image}(d), (e), and (f)).
While our subscript nomenclature is not an \textit{a priori} assumption,
it describes the observed spectral features,
including the sub-peak structure for the northward wind (Fig.\,\ref{fig:lspec}(b)).
Good C-stat/dof values are derived on the \ion{O}{VIII} line (Fig.\,\ref{fig:ofit}(a)).
This model yields similar temperatures with the CIE modelling (Fig.\,\ref{fig:ofit}(b)).
The $\textup{EM}_\textup{S}$/$\textup{EM}_\textup{N}$ ratios
are greater than unity for all position angles (Fig.\,\ref{fig:ofit}(c)):
the brighter $\texttt{vapec}_\textup{S}$ is shifted to the harder side
and the softer side when the dispersion angle is inverted (Fig.\,\ref{fig:lspec}(b)).
This variation naturally arises due to minor differences among RGS slices.
Thus, our shifted-CIEs method approximates successfully and reasonably
the spatial variation and broadening of the \ion{O}{VIII}~Ly$\alpha$ line; that is,
the double CIE modelling with the broadband image can be a surrogate for
an actual emission plasma broadened as the \ion{O}{VIII} band image.

\subsection{Results of \ion{O}{VII} triplet lines}
\label{subsec:ovii}

For the \ion{O}{VII} triplet feature, we also start with the CIE modelling
with the \ion{O}{VII} band image: $\textup{ISM}_\textup{M82} = \texttt{vapec}$.
The ignorance of the point-source contribution aligns with Sect.\,\ref{subsec:oviii}.
On the same philosophy of the \ion{O}{VIII}~Ly$\alpha$ band analysis,
we utilise the 0.45--0.62\,keV band, including the \ion{N}{VII} line at 0.5\,keV.
The spectra are fitted with free parameters of $kT$, emission measure (EM),
and the N and O abundances.
As shown in Fig.\,\ref{fig:lspec}(c), the best-fit CIE models,
convolved with the \ion{O}{VII} band image,
closely resemble the observed line profile of \ion{O}{VII}.
We obtain good C-stat/dof for all dispersions (the $x$-axes of Fig.\,\ref{fig:ofit}(d)).
A hint of residuals is seen at the forbidden line energy of 0.56\, keV,
although their significance is relatively tiny.
For example, in the 319$^\circ$ angle, which exhibits the largest residual, it remains at $\sim 3\sigma$.
The CIE modelling for \ion{O}{VII} provides a uniform $kT_\textup{CIE}$
across the position angles as does the \ion{O}{VIII} band one (Fig.\,\ref{fig:ofit}(e)).
The median value and 16--84th percentile range for $kT_\textup{CIE}$
are 0.14\,keV and 0.11--0.18\,keV, respectively.

Next, we apply the shifted-CIEs modelling for \ion{O}{VII} taking into account its marginal bimodality:
$\textup{ISM}_\textup{M82} = \texttt{vapec}_\textup{N} + \texttt{vapec}_\textup{S}$.
Each CIE component is convolved with the broad RGS bandpass (0.45--1.75\,keV).
Considering the relatively widespread \ion{O}{VII} profile (Figs.\,\ref{fig:image}(d), (e), and (f)),
the additional broadening option of the \textsc{apec} model is adopted.
We set the temperature of the two CIE components to $kT = 0.14$\,keV
based on the results of the CIE modelling.
This model effectively fits the broadened \ion{O}{VII} lines (Figs.\,\ref{fig:lspec}(d)),
yielding C-stat/dof values similar to those from the CIE modellings (Fig.\,\ref{fig:ofit}(d))
The significance of the residuals at 0.56 keV decreased to a maximum of $2\sigma$.
For certain datasets, including that for the position angle of 296$^\circ$,
a single CIE component is sufficient to represent the spectra,
where the dominant component likely corresponds to the bright southwest outflow.

The hint of residuals at 0.56 keV might be attributable to the CX emission.
To constrain the contribution of the CX component,
we employ the second version of the AtomDB CX Model
\textsc{acx} \citep[][]{Smith14}, together with a CIE plasma:
$\textup{ISM}_\textup{M82} = \texttt{vapec} + \texttt{vacx}$ (the CIE+CX modelling).
Both components are convolved in this model with the \ion{O}{VII} band image.
The collision velocity between ions and atoms is set
to moderate values 200\,km\,s$^{-1}$ \citep[e.g.,][]{Cumbee16, Zhang18}.
We assumed that the CIE and CX components share the same temperature (and abundance),
testing two temperature assumptions of $kT =$\,0.1 and 0.2\,keV.
The CIE+CX model also gives fine-fits for all spectra under both temperature assumptions
(Figs.\,\ref{fig:lspec}(e) and \ref{fig:ofit}(d)).
However, this CIE+CX modelling still resulted in similar residuals
around the forbidden line at a level of $\sim 2\sigma$ at most.
Additionally, Fig.\,\ref{fig:ofit}(g) shows that at certain position angles, the CX emission 
significantly exceeds the CIE contribution, reaching up to 
96 per cent at 319$^\circ$, \textit{AND VICE VERSA}.
There is no clear correlation between the CX fraction and position angle;
for example, similar angles yield significantly different CX fractions.
The lack of a consistent pattern leads to the conclusion
that the CIE+CX models may not be physically plausible.

The minor residuals observed at 0.56 keV correspond to the energy
of the forbidden line of \ion{O}{VII} for the bright southward wind component.
We attempted to introduce a spatially narrower CX component to fill the residuals.
We modified the CIE+CX model by replacing the broader CX component
with a more localised and less broadened one.
This modified model does not change our results nor improve spectral fits,
making it challenging to assert the robust presence of the CX emission based on the current RGS data.
\citet{Okon24} recently discussed a possible localised CX emission.
However, they introduce a different model in which the low-energy side
of the observed spectrum is attributed predominantly to the CX emission.

\subsection{Results of broadband spectra}
\label{subsec:broad}

\begin{figure*}
\centering
\includegraphics[width=0.85\linewidth]{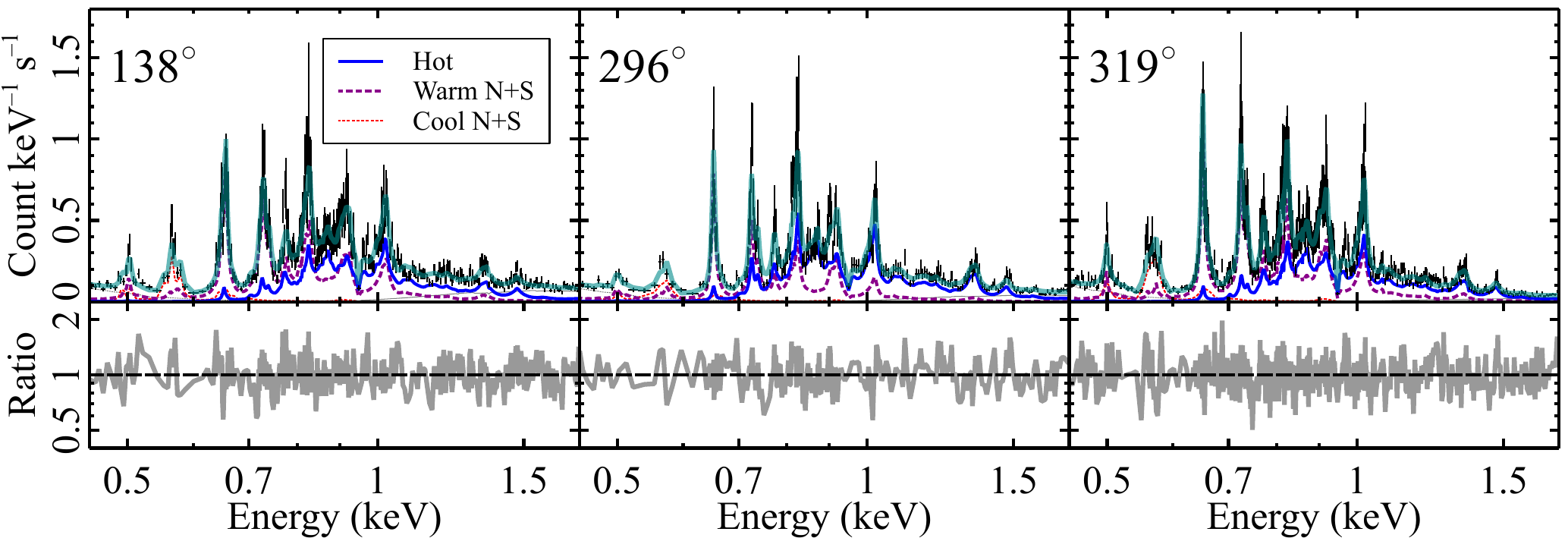}
\caption{The best-fitting models to the broadband RGS spectra
of the M82 centre for the three position angles.
Thin dotted and solid lines represent the background and point-source components,
respectively (see Sect.\,\ref{subsec:fitting}).
\label{fig:bspec}}
\end{figure*}

\begin{figure*}
\centering
\includegraphics[width=0.85\linewidth]{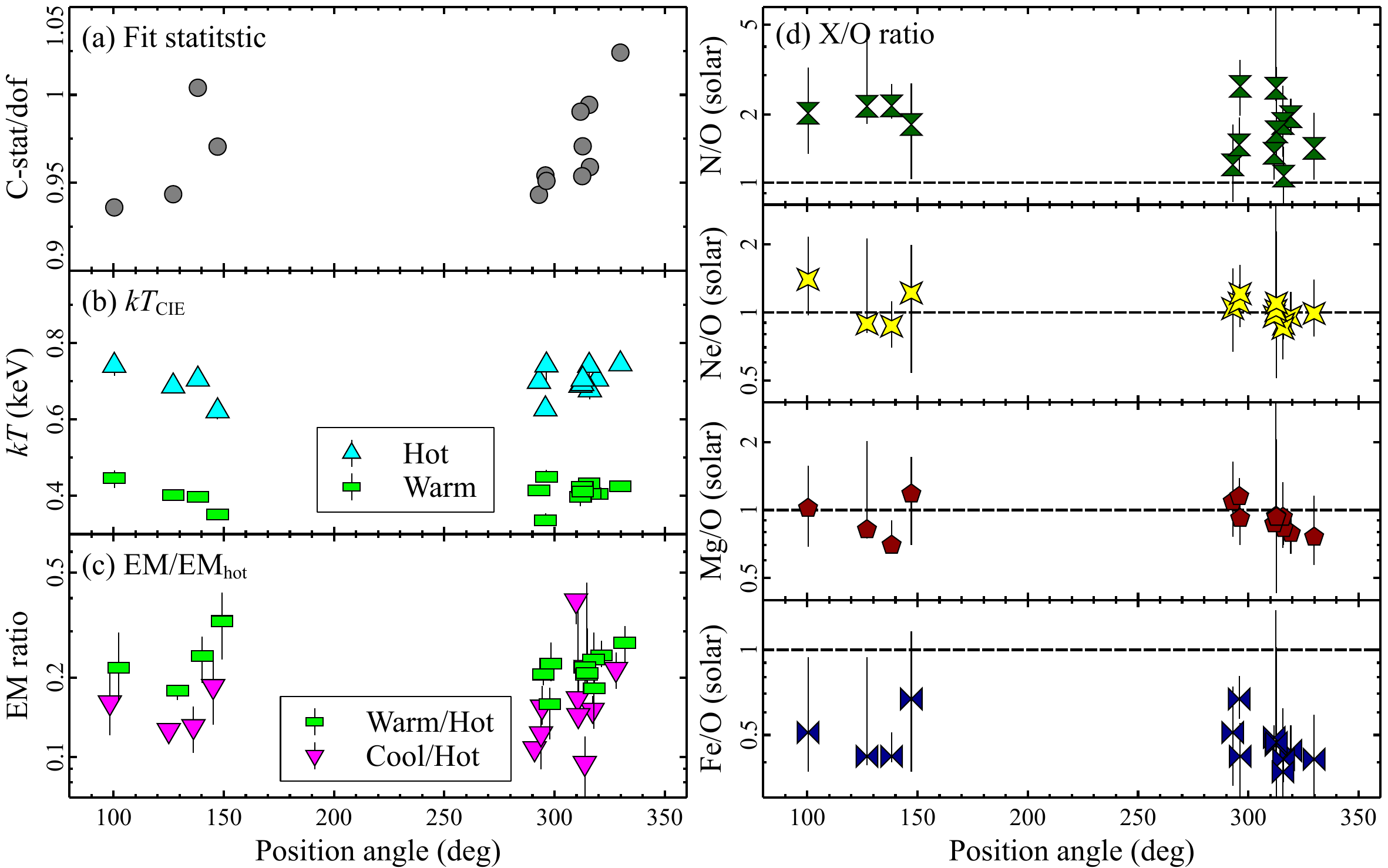}
\caption{(a) C-stat/dof values from the the 0.45--1.75\,keV band fitting
plotted against the RGS position angle.
(b) Temperatures of the warm and hot components.
(c) Relative EM/EM$_\textup{hot}$ ratios for the cool and warm plasmas.
(d) Abundance ratios of N/O (vertical ties), Ne/O (ninja-stars), Mg/O (pentagons), and Fe/O (horizontal ties).
\label{fig:bfit}}
\end{figure*}

Finally, we fit the broadband RGS spectra (0.45--1.75\,keV),
excluding the CX model from this analysis.
Our adopted model is
$\textup{ISM}_\textup{M82} = \texttt{vapec}_\textup{hot}
+ \texttt{vapec}_\textup{warmN} + \texttt{vapec}_\textup{warmS}
+ \texttt{vapec}_\textup{coolN} + \texttt{vapec}_\textup{coolS}$.
All of the CIE components are convolved with the broadband image.
As described in Sects.\,\ref{subsec:oviii} and \ref{subsec:ovii},
two CIEs for the cool and warm plasma, when convolved with the broadband image,
effectively reproduce the local band spectra.
These gas components mainly contribute to the \ion{O}{VII} and \ion{O}{VIII} lines,
with the new hot component mainly accounting for the Fe-L and Mg lines.
In this model, $kT_\textup{hot}$, $kT_\textup{warm}$ and EMs of each component
are varied freely, under the assumption that $kT_\textup{warmN}=kT_\textup{warmS}$.
We fix $kT_\textup{cool}$ to 0.14\,keV.
The N, O, Ne, Mg, Fe, and Ni abundances are free parameters
with shared values across all spectral components, and other metals are set to 1\,solar.
Unfortunately, firm constraints on the Ni abundance are not very promising
since clear Ni line structures are rarely detected by RGS,
even in highly metal-rich systems (e.g., the Centaurus cluster, \citealt{Sanders08, Fukushima22}).
The intrinsic absorption for the hot plasma in M82,
which is more substantial than for the warm and cool gases
(see Sect.\,\ref{subsec:fitting}), is difficult to constrain.
Hence, we adopt $8 \times 10^{21}$\,cm$^{-2}$ for the hot component extinction,
consistent with the intrinsic absorption estimated for the hot gases
by the CCD study ($kT \gtrsim 0.7$\,keV, \citealt{Lopez20}).
Other prescriptions for the warm and cool components follow the ones
with each shifted-CIEs modelling (Sects.\,\ref{subsec:oviii} and \ref{subsec:ovii}).
Despite a bit simplistic modelling, the mean values of the cool+warm and hot gas absorptions
are close to fixed values when allowed to vary.

The best-fitting models for the representative broadband spectra
of M82 are shown in Fig.\,\ref{fig:bspec}.
As we designed, each CIE component contributes to observed emission lines.
This model yields good fits with excellent C-stat/dof values
up to 1.1 for all observations (Fig.\,\ref{fig:bfit}(a)).
In Fig.\,\ref{fig:bfit}(b), we plot the derived $kT_\textup{hot}$
and $kT_\textup{warm}$ against the position angles.
We obtain the medians and 16--84th percentile ranges
as $kT_\textup{hot} = 0.70$\,(0.68--0.74)\,keV
and $kT_\textup{warm} = 0.41$\,(0.40--0.43)\,keV.
The inclusion of the hot gas in the Fe-L lines results
in a slight decrease in the $kT_\textup{warm}$ value compared to
the results from the shifted-CIEs method for the local \ion{O}{VIII} band.
Figure\,\ref{fig:bfit}(c) reveals uniform values for the relative EMs of each temperature plasma.
Here, the hot component is more dominant than the other two, with
the warm and cool gases in similar intensities.
The abundance ratios, N/O, Ne/O, Mg/O, and Fe/O, are also uniform across the dataset:
super-solar N/O, solar ratios of Ne/O and Mg/O, and sub-solar Fe/O (Fig.\ref{fig:bfit}(d)).
The Ni values are not plotted due to large errors and scatters as warned above.
In particular, the solar abundance ratios of light $\alpha$-elements near the disc are the first report,
which will be discussed more in detail in Sect.\,\ref{subsec:abund}.

\section{Discussion}
\label{sec:discussion}

\subsection{Multi-temperature phase in the M82 centre}
\label{subsec:multemp}

In our analysis of the RGS spectra from multiple position angles, we determined that
the hot (0.7 keV), warm (0.4 keV), and cool (0.1--0.2 keV) components effectively 
represent the temperature structure of the ISM in M82.
The 0.6--0.7 keV component, responsible
for Mg (and possibly Si) emission lines, was also reported in studies
using RGS and CCD data \citealt{RS02, Origlia04, Konami11, Zhang14b, Lopez20}).
A warm plasma component with $kT \sim$\,0.5\,keV
has also been identified by \citet{Ranalli08}.
The presence of another cool gas contribution was predicted
through line diagnostics of the \ion{O}{VII} triplet,
suggesting $kT \sim$\,0.1--0.3\,keV, \citealt{Ranalli08}).
However, such a cool component at the M82 core
has not been reported \citep[e.g.,][]{Zhang14b, Lopez20},
except for a hint of the 0.2\,keV component \citep[][but CCD study]{Konami11}. 
The \ion{O}{VII} emission has often been interpreted
as a result of the CX processes \citep[e.g.,][]{Zhang14b}.
However, in our spectral model, which accounts for different spatial broadenings
for \ion{O}{VII} and \ion{O}{VIII}, the CIE components
are the primary contributors to these line emissions.

Observational and simulation studies generally predict the core regions of starburst galaxies
undergo a multi-temperature phase of gas, ranging from cold dust to hot outflows \citep[e.g.,][]{Leroy15}.
Interestingly, such a multiphase state is also applicable to X-ray-emitting gas itself
\citep[e.g.,][]{Melioli13, Schneider18}.
This theoretical prediction is in good agreement with the plasma content that we revealed for M82,
as well as with other starburst galaxies (e.g., NGC~3079, \citealt{Konami12}; Arp~299, \citealt{Mao21}).
Furthermore, even in the Milky Way, certain regions with a high concentration of massive stars,
such as superbubbles, possess intermixed X-ray plasma with different temperatures
(e.g., \citealt{Kim17} for simulation; \citealt{Fuller23} for observation).
The multiphase ISM would be ubiquitous in star-forming regions, regardless of their scale.
While the origin of different spatial variations of the cool and warm components
eludes our present study, we expect it to be examined more robustly
through the high-resolution observation of M82.

\subsection{Note of CX emission}
\label{subsec:cx}

Previous reports on the CX emission from M82 did focus on the limited position angle
(mainly the 319$^\circ$ observation, \citealt{Liu11, Zhang14b}).
In this angle, the \ion{O}{VII} profile expands toward the low energy scale
(Figs.\,\ref{fig:image}(d) and (f)).
This expansion is interpreted as high R- or G-ratio values
\footnote{These ratios are given as $f/i$ (R) and $(i+f)/r$ (G),
where $r$, $f$, and $i$ are the flux of resonance, forbidden,
and intercombination lines, respectively} in the energy space.
The enhancement in the soft part of the \ion{O}{VII} triplet is likely 
due to the bright outflow towards the southeast, as discussed in Sect.\,\ref{subsec:ovii}.
While our current results do not strongly favour CX emissions,
we consider its contribution in M82 using the 319$^\circ$ result that is
the most significantly contaminated by the CX emission if any.
We assume that the CX and CIE components share the same spatial distribution,
although a more localised CX emission can be debated (Sect.\,\ref{subsec:ovii}).
From this observation, we obtained an \textsc{acx} normalisation of $(3.3\pm 1.0) \times 10^{-4}$\,cm$^{-5}$.
According to the study of M51 by \citet{Yang20},
we can evaluate the spatial scale of the CX reaction from this \textsc{acx} normalisation as
\begin{equation}
\label{eq:1}
\frac{10^{-10}}{4\pi D^2} \int \frac{n}{\sigma}\,\textup{d}A\textup{.}
\end{equation}
Here, $D$, $n$, and $\sigma$ are the angular diameter distance to M82,
the density of receiver ions in the CX reaction and the CX cross sections, respectively.
Assuming a spherical volume with a 40\,arcsec ($\sim$\,0.67\,kpc at M82) radius,
we adopted $n = 0.19$\,cm$^{-3}$ from the CIE modelling in Sect.\,\ref{subsec:ovii}.
The cross sections were set as $\sigma = 5\times 10^{-15}$\,cm$^{2}$
\citep[][and references therein]{GS23}.
In consequence, we obtained the CX emitting area $A = 14\pm 3$\,kpc$^2$,
Similar illustrations are reported by \citet{Zhang14b} for the same M82 cente and \citet{Yang20} for M51.
This significant discrepancy makes us suspicious about the brightness of the CX emission
if the broadened CX component is a dominant source of the \ion{O}{VII} triplet lines.
Indeed, \citet{Okon24} propose that the CX component accounts for 40--60\,per cent
of the \ion{O}{VII} flux using a multi-temperature model,
substantially lower than the previous estimate ($\sim$\,90\,per cent, \citealt{Zhang14b}).
The CX emission properties in M82, such as luminosity or spatial extent,
must be under continuous vigilance to be challenged or validated more confidently
with the high-resolution and non-dispersive X-ray spectroscopic data
coming down from \textit{XRISM} in orbit.

\subsection{Metal abundance pattern}
\label{subsec:abund}

\begin{figure}
\centering
\includegraphics[width=0.85\linewidth]{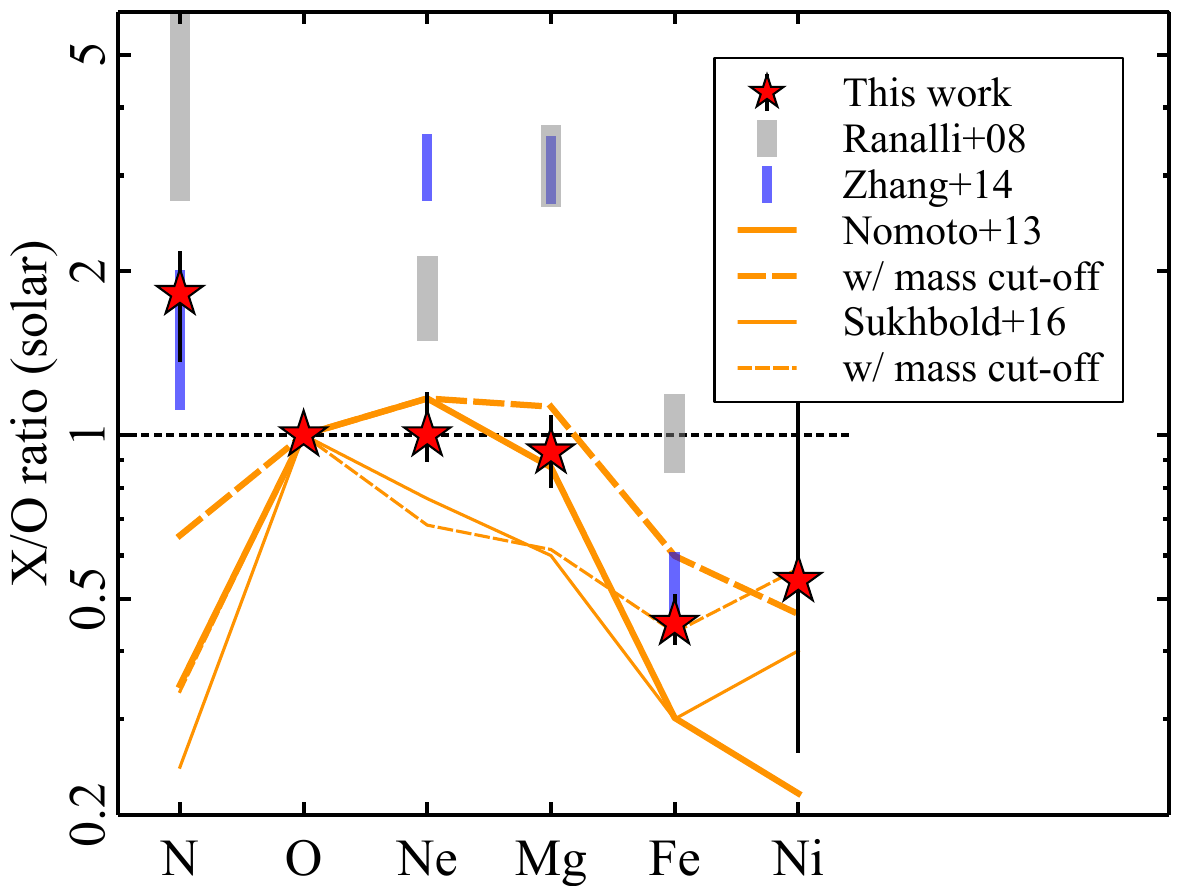}
\caption{Observed X/O abundance pattern from the broadband spectral fits,
where the medians and 16--84th percentile ranges for all dispersion angles are plotted.
Two literal values by previous RGS works are given for comparison.
The dashed line represents the solar composition.
The thick and thin solid lines indicate the IMF-weighted yields predicted
by \citet{Nomoto13} and \citet{Sukhbold16}, respectively,
assuming the solar initial metallicity for progenitors.
The dashed lines are for each option of the mass-cutoff integration
($< 25$\,$\textup{M}_{\odot}$).
\label{fig:abund}}
\end{figure}

The metal abundance pattern is a key to studying
the chemical enrichment process of observed gases.
Figure\,\ref{fig:abund} shows the median values and the 16--84th percentile ranges
of N/O, Ne/O, Mg/O, and Fe/O across all position angles.
For comparison, we also plot the Ni/O value, which is in possible agreement with Fe/O within a large error.
We obtained solar-like ratios of the Ne/O and Mg/O
that are lower than the results of two earlier RGS works \citep[][]{Ranalli08, Zhang14b}.
Other studies with CCD reported super-solar patterns of light $\alpha$-elements \citep[e.g.,][]{Konami11}.
The presence or absence of the CX contribution does not change
the super-solar abundances in these studies.
The primary origin of these gaps would be our careful modelling
of the \ion{O}{VII} and \ion{O}{VIII} lines rather than the presence of CX emission,
which improves the estimation of a temperature structure in M82.
In particular, the inclusion of the cool gas component would be essential progress
in measuring the element abundances.

In Fig\,\ref{fig:abund}, we compare the observed abundance pattern 
in the M82 centre, a starburst system, with nucleosynthesis models of CCSNe.
This comparison is crucial as most supernovae
in such systems are expected to originate from massive stars. 
We adopt two models: the ``classical'' standard yield by \citet{Nomoto13}
and the latest N20 calculations by \citet{Sukhbold16}.
For both models, we assume an initial mass function (IMF) by \citet{Salpeter55} 
with progenitor masses up to 40\,$\textup{M}_\odot$.
In addition, based on the discussion on the missing progenitor mass problem
for CCSNe by \citet[][]{Smartt15}, we also employ another IMF-weighted integration
with the upper mass limit of 25\,$\textup{M}_\odot$.

For $\alpha$-elements, the yields with \citet{Nomoto13} most closely match
the observed Ne/O and Mg/O ratios but tend to underestimate the Fe/O and Ni/O ones.
For the case with the upper mass limit of 25\,$\textup{M}_\odot$,
the half-solar Fe/O and Ni/O values are explained well
by both of \citet{Nomoto13} and \citet{Sukhbold16}.
In a powerful starburst galaxy Arp~299, \citet{Mao21} provide a similar picture
by comparing the model yields to the N/O, Ne/O, Mg/O, S/O, and Ni/O pattern.
However, the observed Fe/O and Ni/O ratios in M82 might also be
explained by including a minor contribution from Type~Ia SNe
(only 6 per cent of the total SNe in the M82 core).
The hot ISM in M82, as in other starburst galaxies,
serves as a significant repository of CCSN products.

None of these CCSNe models adequately reproduce the N/O ratio $\sim$\,2\,solar in M82.
Similar high N/O ratios have been reported in more ancient systems,
such as early-type galaxies, including the brightest cluster galaxies \cite[e.g.,][]{Mao19, Fukushima23a}.
These objects owe a dominant fraction of their N enrichment to mass-loss winds
from massive and asymptotic giant branch stars.
Thus, such mass-loss channels, aside from CCSNe,
likely contribute to the enrichment observed in the core of M82.
One caveat is that the \ion{N}{VII}~Ly$\alpha$ line is present at $\sim$\,0.5\,keV
and is likely to be affected by, for example,
the estimation of the absorption for the cool+warm component
and/or the contribution of the background emission (Sect.\,\ref{sec:result}).
When the gate valve is opened, the \textit{XRISM} data will provide us
a great opportunity to constrain the N abundance more robustly.

\begin{acknowledgements}
The authors would like to thank Dr.~D.~Wang for taking the time to read our manuscript thoroughly
as a reviewer and for providing helpful comments and suggestions.
K.F. shall deem it an honour to be supported by the Japan Society for the Promotion of Science (JSPS)
through Grants-in-Aid for Scientific Research (KAKENHI)
grant Nos.~21J21541 and 22KJ2797 (Grant-in-Aid for JSPS Fellows).
This work is based on observations obtained with \textit{XMM-Newton},
an ESA science mission with instruments and contributions
directly funded by ESA Member States and NASA of the USA.
The \textit{XMM-Newton} Science Archive (\url{https://nxsa.esac.esa.int/nxsa-web/})
stores and distributes the RGS data analysed in this paper.
We get the second version of \textsc{acx} from the \textsc{AtomDB} website (\url{http://www.atomdb.org/CX/})
and run it on the \textsc{pyxspec} module implemented in the \textsc{xspec} package.
The figures in this paper are generated using \textsc{veusz}
(\url{https://veusz.github.io}) and \textsc{python} (\url{https://www.python.org}).
\end{acknowledgements}

%
%

\bibliographystyle{aa}
\bibliography{paper_ref}

\end{document}